\documentclass[manuscript]{acmart}

\AtBeginDocument{%
  \providecommand\BibTeX{{%
    \normalfont B\kern-0.5em{\scshape i\kern-0.25em b}\kern-0.8em\TeX}}}

\copyrightyear{2022}
\acmYear{2022}
\setcopyright{rightsretained}
\acmConference[RecSys '22]{Sixteenth ACM Conference on Recommender Systems}{September 18--23, 2022}{Seattle, W A, USA}
   
\acmBooktitle{Sixteenth ACM Conference on Recommender Systems (RecSys '22), September 18--23, 2022, Seattle, WA, USA} \acmDOI{10.1145/3523227.3551467} 
\acmISBN{978-1-4503-9278-5/22/09}

\usepackage{booktabs} 
\usepackage{color}

\begin{document}

\fancyhead{}
\title{DAGFiNN: A Conversational Conference Assistant} 

\author{Ivica Kostric}
\orcid{0000-0002-0859-9762}
\affiliation{%
  \institution{University of Stavanger}
  \city{Stavanger}
  \country{Norway}
}
\email{ivica.kostric@uis.no}
\author{Krisztian Balog}
\affiliation{%
  \institution{University of Stavanger}
  \city{Stavanger}
  \country{Norway}
}
\email{krisztian.balog@uis.no}
\author{Tølløv Alexander Aresvik}
\affiliation{%
  \institution{University of Stavanger}
  \city{Stavanger}
  \country{Norway}
}
\email{ta.aresvik@stud.uis.no}
\author{Nolwenn Bernard}
\affiliation{%
  \institution{University of Stavanger}
  \city{Stavanger}
  \country{Norway}
}
\email{nolwenn.m.bernard@uis.no}
\author{Eyvinn Thu Dørheim}
\affiliation{%
  \institution{University of Stavanger}
  \city{Stavanger}
  \country{Norway}
}
\email{et.dorheim@stud.uis.no}
\author{Pholit Hantula}
\affiliation{%
  \institution{University of Stavanger}
  \city{Stavanger}
  \country{Norway}
}
\email{pf.hantulabergan@stud.uis.no}
\author{Sander Havn-Sørensen}
\affiliation{%
  \institution{University of Stavanger}
  \city{Stavanger}
  \country{Norway}
}
\email{s.havn-sorensen@stud.uis.no}
\author{Rune Henriksen}
\affiliation{%
  \institution{University of Stavanger}
  \city{Stavanger}
  \country{Norway}
}
\email{r.henriksen@stud.uis.no}
\author{Hengameh Hosseini}
\affiliation{%
  \institution{University of Stavanger}
  \city{Stavanger}
  \country{Norway}
}
\email{h.hosseini@stud.uis.no}
\author{Ekaterina Khlybova}
\affiliation{%
  \institution{University of Stavanger}
  \city{Stavanger}
  \country{Norway}
}
\email{e.khlybova@stud.uis.no}
\author{Weronika Lajewska}
\orcid{0000-0003-2765-2394}
\affiliation{%
  \institution{University of Stavanger}
  \city{Stavanger}
  \country{Norway}
}
\email{weronika.lajewska@uis.no}
\author{Sindre Ekrheim Mosand}
\affiliation{%
  \institution{University of Stavanger}
  \city{Stavanger}
  \country{Norway}
}
\email{se.mosand@stud.uis.no}
\author{Narmin Orujova}
\email{n.orujova@stud.uis.no}
\affiliation{%
  \institution{University of Stavanger}
  \city{Stavanger}
  \country{Norway}
}

\begin{abstract}
DAGFiNN is a conversational conference assistant that can be made available for a given conference both as a chatbot on the website and as a Furhat robot physically exhibited at the conference venue.
Conference participants can interact with the assistant to get advice on various questions, ranging from where to eat in the city or how to get to the airport to which sessions we recommend them to attend based on the information we have about them.
The overall objective is to provide a personalized and engaging experience and allow users to ask a broad range of questions that naturally arise before and during the conference.
\end{abstract}



\maketitle

\section{Introduction}

Conversational systems for information access have received much attention in recent years~\citep{anand:2020:dagstuhl,gao:2019:FnTIR}, with conversational recommendation being one of the main tasks~\citep{Hauff:2021:TOIS, Gao:2021:AIO}.
A conversational recommender system (CRS) is a multi-turn, interactive system that can elicit user preferences in real-time using natural language~\citep{Jannach:2021:ACMCS}.
Common research threads related to CRS include question-based user preference elicitation~\citep{Kostric:2021:RecSys}, multi-turn conversational recommendation strategies~\citep{Lei:2020:WSDM}, natural language understanding and generation~\citep{Li:2018:NeurIPS}, and evaluation and user simulation~\citep{Zhang:2020:KDD}. With some exceptions~\citep{Walker:2004:CogSci, Arteaga:2019:CISTI}, the vast majority of work uses text as the sole modality and targets a single domain. Arguably, in many real-world contexts, a rich user experience would encompass multiple modalities and domains.

Specifically, in this paper we study the problem of conversational recommendations in the context of an academic conference. 
To mimic the capabilities of a person sitting behind the information desk, a CRS needs to have basic conversational skills as well as knowledge about things that conference participants would typically ask about, such as recommendations for places to eat and to visit, and conference schedule.
A person behind the counter would also personalize responses, remember past interactions, and use additional digital/printed material to help convey information (e.g., show directions to a restaurant on a map).
The system we present here, DAGFiNN, represents a step towards the digital version of such conference assistant.

\begin{figure}[t]
    \centering%
    \includegraphics[width=.3\linewidth]{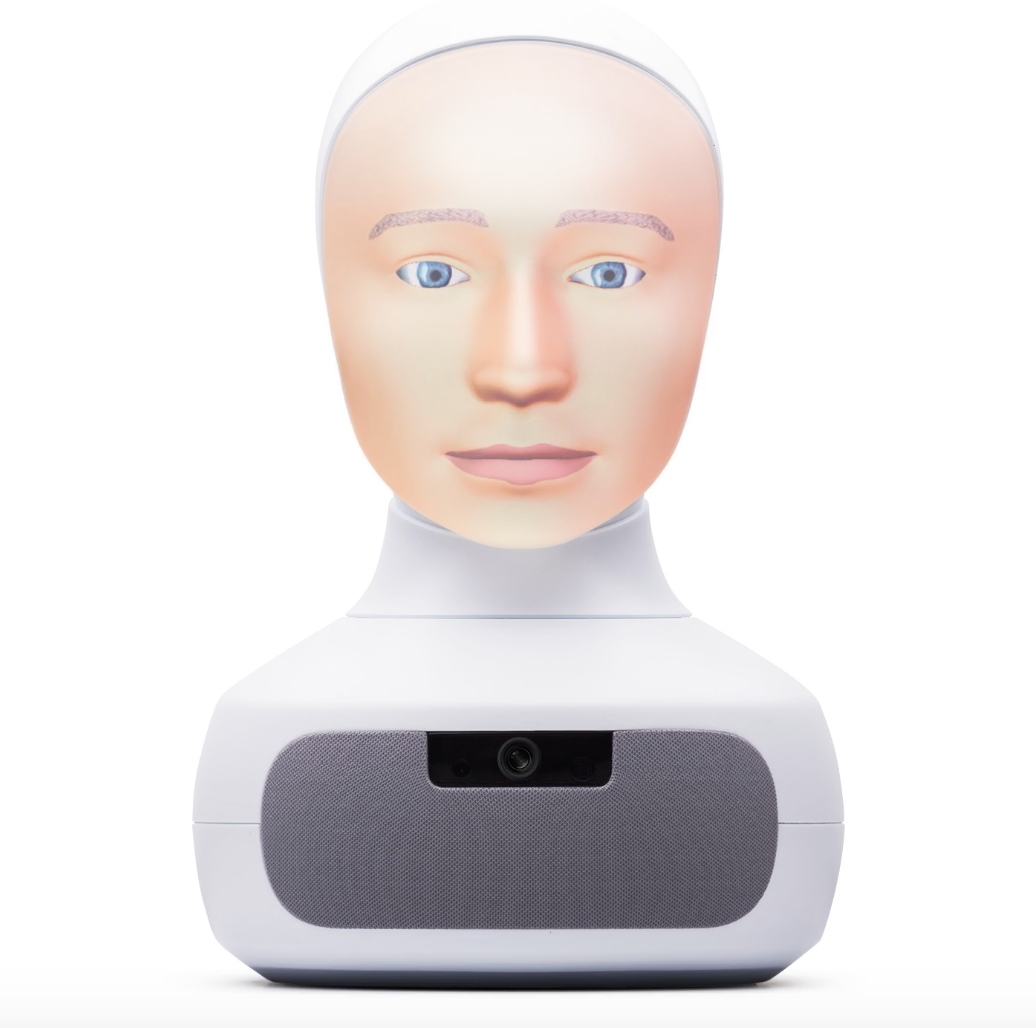}
    \includegraphics[width=.69\linewidth]{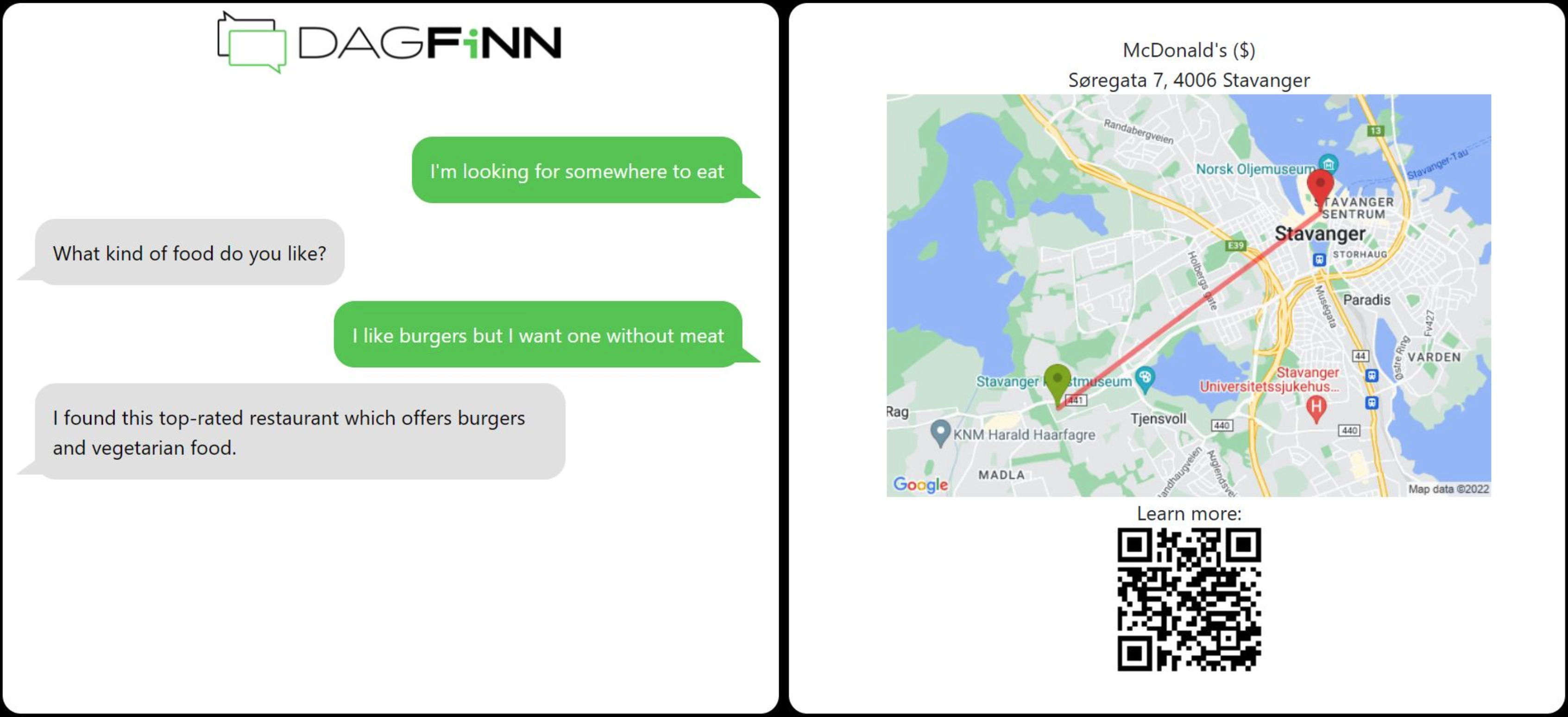}
    \caption{Furhat (Left) with the external screen (Right), showing an excerpt from a POI recommendation dialogue.}
    \Description[]{Furhat (Left) with the external screen (Right), showing an excerpt from a POI recommendation dialogue.}
    \label{fig:func}
\end{figure}

To the best of our knowledge, DAGFiNN is a first of its kind, being an open-source, multi-domain, multi-modal conversational recommender system. The specific research challenges we address are as follows:

\begin{itemize}
    \item Provide a holistic user experience by providing recommendations across multiple domains, basic conversational capabilities, and personalization integrated into a single system.  Currently, recommenders are built for point-of-interest (POI) and conference content, but the set of domains can be extended in a modular fashion.
    \item Support multiple user interfaces (UI), specifically, a Web chat and a Furhat robot with an optional external screen, and use rich responses customized to each user interface to effectively convey information.
    \item Utilize multiple input modalities for the Furhat robot. Furhat is a social robot with a microphone, speaker, and camera, which makes it possible to recognize users that interacted with the robot previously.
\end{itemize}
\noindent
The DAGFiNN project is the result of the combined effort of multiple student teams, each responsible for a specific component.
DAGFiNN has been field-tested at an international conference, which we briefly report on in Section~\ref{sec:experiment}.
The source code, more detailed descriptions of the skills, and a demonstration video are available at \url{https://dagfinn.ai}.
\section{Related work}
\label{sec:related}

Recommender systems have been extensively studied for the academic~\citep{Gingstad:2020:CIKM} and point-of-interest (POI)~\citep{sanchez:2022:ACMCS,li:2021:DASFAA} domains.  
There is a rich body of recommendation methods to build on.  However, our focus in this work is on the conversational user experience; more advanced recommendation methods may be incorporated in the future. 
Notably, the type of items that can be recommended is slightly different in our case. For example, the majority of the research on academic recommendation focuses on scientific literature recommendation~\citep{Li:2018:IEEE, bai:2019:IEEE,Gingstad:2020:CIKM}, while in our work, similarly to \citep{Yi:2018:IEEE}, the focus is on recommending conference sessions.
We focus not only on recommending POIs, but also on how to get there.

There are currently only a few open-source CRS prototypes available.
CRSLab~\citep{zhou:2021:IJCNLP} and Macaw~\citep{Zamani:2020:SIGIR} are general frameworks, while VoteGoat~\citep{Dalton:2018:SIGIR} and IAI MovieBot~\citep{habib:2020:CIKM} are specific to the movies  domain.  None of these systems has support for rich multi-modal interactions across multiple user interfaces.

\section{Conversational Conference Assistant}
\label{sec:func}

DAGFiNN is a conversational conference assistant that can be made available both as a chatbot on the conference website and as a Furhat robot physically exhibited at the conference venue.
Conference participants can interact with DAGFiNN to get advice on a variety of questions, from point-of-interest (POI) recommendations to suggestions on which sessions to attend at the conference.
DAGFiNN has several skills, modular in design, making it easy to add new ones or remove the ones that are no longer needed. As its \textbf{Core} skill, it is equipped with basic ``chit-chat'' conversational capabilities, to handle questions like \emph{Who are you?} or \emph{What's the weather like in Seattle?}
It uses simple rules to determine the appropriate response for some anticipated questions. These feature relies mostly on templated responses to frequently asked queries and handles out-of-scope utterances.

In order to provide a personalized experience, there is an additional personalization layer to allow for tailoring responses to individuals. 
In its current form, DAGFiNN can recognize users interacting with the Furhat robot by scanning a QR code from their badge or, if permitted, by taking pictures of users' faces during their interaction.  

In terms of recommendation skills, we target two domains that represent some of the most common information needs that arise in a conference context: POI and conference content.
DAGFiNN can elicit user preferences via multi-turn interaction to narrow the search space and can handle follow-up questions about items.
The \textbf{POI recommender} answers queries about restaurants, museums, and other activities in the vicinity of the conference venue, like \emph{Where can I go running?} or \emph{Do you know of any good Indian restaurants?} Once the user has made a selection, they can ask follow-up questions, e.g., \emph{How do I get there?} 
The collection of items is extracted from TripAdvisor and Google into a manually curated database, with details on name, category (e.g., restaurant, bar, museum, park), price range, address, rating, etc.
The skill is triggered by asking for a recommendation on one of the categories. After a preference elicitation step, where preferences are represented internally as lists of liked and disliked keywords, a recommendation is made on each turn from the set of top-rated matching items.
When the user is satisfied with the recommendation, they can further ask questions about the recommended item or inquire about transport options.

The \textbf{conference content recommender} skill can provide details on the conference programme (keynotes, tutorials, workshops, sessions) and authors, and can give recommendations on which session to attend based on elicited interests. Example queries include \emph{Who are the conference's keynote speakers?}, \emph{Can you recommend me a session to attend?} and \emph{What is the next session about?}
The items (sessions, speaker, schedule, and room information) are obtained from the conference website.

\section{Architecture}
\label{sec:appr}

\begin{figure*}[t]
    \centering
    \includegraphics[width=0.95\linewidth]{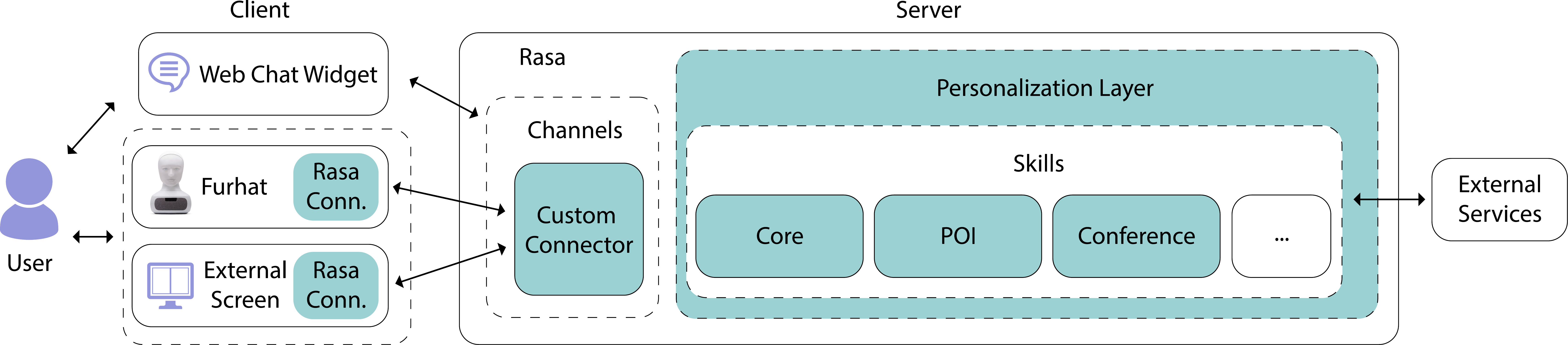}
    \caption{DAGFiNN architecture. Left: Clients with different interfaces; Web chat and Furhat robot with external screen. Right: Server.}
    \Description[]{DAGFiNN architecture. Left: Clients with different interfaces; Web chat and Furhat robot with external screen. Right: Server.}
    \label{fig:architecture}
\end{figure*}

DAGFiNN follows a client-server architecture; the main components are shown in Figure~\ref{fig:architecture}.
The server is built on top of the open-source Rasa conversational framework.\footnote{\url{https://rasa.com/open-source/}} 
Rasa handles dialogue management by classifying intents and then invoking the respective skills to generate a response. Skills are implemented using a combination of rules, stories, and forms, which represent the various (increasingly complex) ways to provide training data in Rasa to train the dialogue management model.
We implemented a logging system, which stores all conversations in a database for later analysis.
Rasa also features connector classes to many different messaging platforms (e.g., Slack, Telegram). The notion of a \emph{channel} in Rasa refers to a communication path between a server and a client (e.g., Rest API, socket.io), on top of which we can develop custom features for specific platforms. 

DAGFiNN supports two types of user interfaces at the moment: a Web chat and Furhat robot with an optional external screen.
A JavaScript Web chat widget is provided out of the box by Rasa that can be added to any webpage. 
The Furhat robot talks to the server using a custom Furhat `skill' written in Kotlin and using the socket.io library. Similarly, the external screen uses the same socket.io channel to connect to the same custom connector in Rasa server, thereby the conversation is synchronized between the two devices.

Responses may be customized for different channels, to fit the modalities available on that channel. 
Finally, the personalization layer handles the identification of users based on facial recognition or QR code, displayed on the participant's badge. Before taking pictures, for compliance with GDPR, DAGFiNN always asks permission to try to identify users.  Knowing the user's identity allows for the development of additional features in the future. 

\section{Field Study}
\label{sec:experiment}

\begin{figure}[t]
    \centering%
    \includegraphics[width=.53\linewidth]{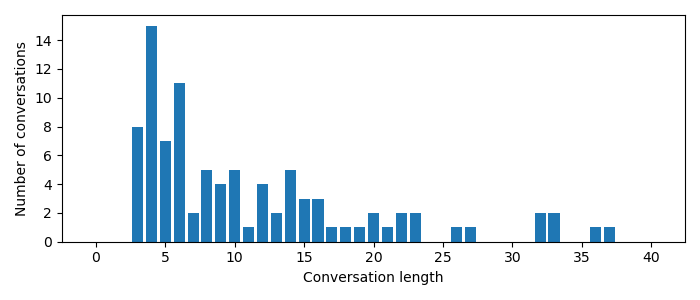}
    \includegraphics[width=.379\linewidth]{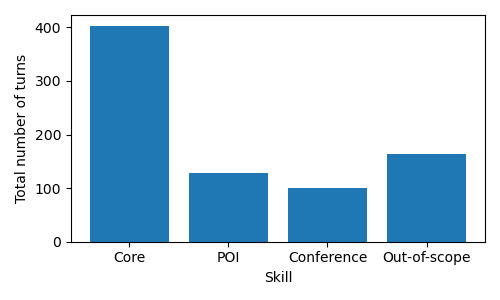}
    \caption{Left: Distributions of conversation lengths. Right: Distributions of turns per skill.}
    \Description[]{Left: Distributions of conversation lengths. Right: Distributions of turns per skill.}
    \label{fig:experiment}
\end{figure}

DAGFiNN has been featured as a digital receptionist at the ECIR’22 international conference, where participants could interact with it to get recommendations related to places to visit (POIs) and to the content of the conference.  It has offered an engaging experience, especially for POI recommendations, and has received overwhelmingly positive feedback.
Figure~\ref{fig:experiment} (Left) shows the distribution of conversation lengths; we observe that most conversations last 3-6 turns. On the right, we show the distribution of intents across skills. Unsurprisingly, the most commonly triggered skill was Core, as most conversations include intents like greetings and good-byes, which are a part of the Core skill. 

While users in general liked the recommendations they received, some were probing the abilities with more perplexing requests like \emph{Where to go for a swim?} or \emph{I am interested in seeing the fjords.} 
Also, given the Furhat robot interface, some people seemed to be more interested in a social chitchat-like experience than inquiries about actual information needs.

\section{Conclusion and Future Work}
\label{sec:concl}

We have presented DAGFiNN, an open-source conversational conference assistant. It has been developed using the task-oriented open-source framework Rasa, and run on a social Furhat robot along with an external display, allowing for both spoken conversations and visual information. 
DAGFINN has been field-tested at an international conference and has received positive feedback. 
Nevertheless, it is still in a research prototype phase and we plan to expand its functionality in the future. Some ideas include adding new skills, getting implicit feedback through facial expressions, supporting multi-party dialogues, and extending the personalization aspect.
Given that DAGFiNN has been developed by a group of students as part of their bachelor/master theses, we are also interested in utilizing it as an educational tool and as a platform for other student projects in the future.

\bibliographystyle{ACM-Reference-Format}
\bibliography{recsys2022-dagfinn.bib}

\end{document}